\documentclass[preprint,eqsecnum,aps]{revtex4}
\usepackage{graphics}
\usepackage{relsize}
\usepackage{array}
\usepackage{tabularx}
\usepackage{rotating}
\usepackage{relsize}
\usepackage{amssymb}

\def\X{\mbox{\boldmath $X$}}
\def\x{\mbox{\boldmath $x$}}
\def\r{\mbox{\boldmath $r$}}
\def\B{\mbox{\boldmath $B$}}
\def\U{\mbox{\boldmath $U$}}
\def\V{\mbox{\boldmath $V$}}
\def\F{\mbox{\boldmath $F$}}
\def\Y{\mbox{\boldmath $Y$}}
\def\0{\mbox{\boldmath $0$}}
\def\O{\mbox{\boldmath $O$}}
\def\SMAT{\mbox{$\mathbb{S}$}}

\begin{document}

\title{The second-order reduced density matrix method and the two-dimensional Hubbard model}

\author{James S. M. Anderson}
\affiliation{Department of Physics, The University of Tokyo,7-3-1 Hongo, Bunkyo-ku, Tokyo, Japan, 113-0033, Japan}

\author{Maho Nakata}
\email{maho@riken.jp}
\affiliation{Advanced Center for Computing and Communication, RIKEN, 2-1 Hirosawa, Wako-city, Saitama, 351-0198, Japan}

\author{Ryo Igarashi}
\affiliation{Institute for Solid State Physics, The University of Tokyo,
5-1-5 Kashiwanoha, Kashiwa, Chiba, 277-8581, Japan}

\author{Katsuki Fujisawa}
\affiliation{Chuo University and JST CREST, 1-13-27 Kasuga, Bunkyo-ku, Tokyo, 112-8551, Japan}

\author{Makoto Yamashita}
\affiliation{Department of Mathematical and Computing Sciences,
Tokyo Institute of Technology, 2-12-1-W8-29 Ookayama, Meguro-ku, Tokyo 152-8552, Japan.}

\begin{abstract}
The second-order reduced density matrix method (the RDM method) has performed 
well in
determining energies and properties of atomic and molecular systems, achieving coupled-cluster singles and doubles with perturbative triples (CCSD(T)) accuracy without
using the wave-function. 
One question that arises is how well does the RDM method perform with the same conditions that
result in CCSD(T) accuracy in the strong correlation limit. 
The simplest and a theoretically important model for strongly correlated electronic systems is the Hubbard model.
In this paper, we establish the utility of the RDM method when employing the $P$, $Q$, $G$, $T1$ and $T2^\prime$ conditions in the two-dimensional Hubbard model case and we conduct a thorough study applying the $4\times 4$ Hubbard model employing a coefficients. Within the Hubbard Hamiltonian we found that even in the intermediate setting, where $U/t$ is between $4$ and $10$, the $P$, $Q$, $G$, $T1$ and $T2^\prime$ conditions reproduced good ground state energies.
\end{abstract}
\maketitle

\section{Introduction}
The second-order reduced density matrix is necessary and sufficient to compute all the physical properties that one can compute using the wave-function \cite{DM1960}.
Due to its simplicity, it has been a dream for quantum chemists to directly determine the second-order reduced density matrix instead of using the wave-function, and we believe that it should be simpler to determine than solving the Schr\"odinger equation. 

When an appropriate subset of necessary $N$-representability conditions, a term coined by Coleman \cite{Coleman63}, are used as constraints in a variational calculation of the second-order reduced density matrix one is able to compute accurate energies of the second-order reduced matrices producing accurate energies and properties. This approach is known as the RDM method
and has a long history \cite{OldRDM,MihailovicRosina}. Unfortunately, the RDM method faded away because no algorithm for systematic calculations was available at the time and the $N$-representability condition was not very well understood. 

After 25 years, in 2001, Nakata {\it et al.} formulated the RDM method as the standard form of the primal semidefinite programming problem. They performed a systematic study on small (few electron) atoms and molecules \cite{Nakata01}. They used the $P$, $Q$ \cite{Coleman63},and $G$ conditions \cite{Garrod64} as the $N$-representability constraints that resulted in 120\% of correlation energies. These promising results led Zhao {\it et al.} three years later to include the $T1$, and the $T2$-conditions in addition to the $P$, $Q$, and $G$ conditions in the RDM method giving results with similar accuracy to coupled-cluster singles and doubles with perturbative triples (CCSD(T)) for atomic and molecular systems \cite{Erdahl78,Zhao04,Nakata08}. Since then, research along these lines has spread with enthusiasm and several papers have been published \cite{RDMRecent}.

However, the correlation in molecular systems is not especially strong. We want to investigate the robustness of these conditions in predicting accurate energies in the case of strong correlation. To test this, we have chosen to employ the Hubbard model \cite{HUBBARD63}. This model is interesting not only because of its simplicity, but also its capability of describing strong electron correlation. The RDM method has been applied to the Hubbard model by Hammond {\it et al.} \cite{Hammond}, Nakata {\it et al.} \cite{Nakata08}, and Verstichel {\it et al.} \cite{Verstichel}. Their results very accurately described total energies as well as other properties. However, they only treated the one-dimensional Hubbard model, which can be solved analytically by the Bethe-Ansatz as demonstrated by Lieb and Wu \cite{LiebWu}. It can also be treated numerically by the density matrix renormalization group (DMRG) method \cite{DMRG}. As a result, the behavior of the correlation is rather well understood \cite{onedhubbard}. 

 The challenge for the condensed matter physics community is, thus, to compute the ground state energy and properties of the two-dimensional Hubbard model since no analytic results are available as they are in the one-dimensional case. Still, it is an open question, but it is believed that two-dimensional Hubbard model is the simplest model that exhibits the high-$T_c$ superconductivity of copper oxide \cite{Anderson}. The underlying physics of the Hubbard Hamiltonian remains a topic of considerable discussion \cite{HighTc_ornot}. 

This problem can be reduced to the eigenvalue problem of astronomically large symmetric matrices. Extensive numerical studies \cite{Dagotto} have been done using the Quantum Monte Carlo (QMC) method, the Exact Diagonalization (ED) method (also known as the full configuration interaction (FCI) method), and the DMRG method.
 However, we can solve very small two-dimensional Hubbard model system without much difficulty. To the best of the authors' knowledge the largest two-dimensional Hubbard model systems that have been treated are the $10\times 10$ square lattice by Sorella or $16\times 16$ square lattice by Chen {\it et al.} \cite{QMC}, $40$ to $64$ rectangular or square lattices by the DMRG \cite{largedmrg}, and the $\sqrt{20} \times \sqrt{20}$ by the Exact Diagonalization \cite{Tohyama}. Aside from the Exact Diagonalization, the accuracy of the ground state energies can be 
dubious.

Advantages of the RDM method are: this method calculates the lower bound to the FCI energy in the same basis set whereas all of the other methods give upper bounds, thus this approach is complementary to the former methods. This method does not require extrapolation to the absolute zero-temperature. It does not suffer from minus sign problem in QMC \cite{QMCminussign}. It does not depend on the choice of lattice which may appear in DMRG calculation \cite{DMRGchoice}. 

In this paper, we calculated the total energies of the two-dimensional Hubbard model using the RDM method and compared them to the exact results from ALPS \cite{ALPS} to examine whether the $P$, $Q$, $G$, $T1$ and $T2^\prime$ conditions are physically important in strongly correlated system. The rest of paper is organized as follows. In Section II, we briefly review the RDM method, the $N$-representability conditions, semidefinite programming, and the Hubbard models. The results and discussion are shown in Section III. The conclusions are in Section IV.

\section{Theory} 

\subsection{Reduced density matrices}
The second-order density matrix is an example of a broader class of density matrices. The most general form is the $M$-th order density matrix. This has the form
\[
{}^{(M)}\Gamma^{i_1 i_2 \cdots i_M}_{j_1 j_2 \cdots j_M} =\frac{1}{M!}\langle \Psi | a_{i_1}^\dagger
a_{i_2}^\dagger \cdots a_{i_M}^\dagger a_{j_M} \cdots a_{j_2} a_{j_1}  |  \Psi  \rangle
\]
The second-order reduced density matrix is an important special case. This is the reduced density matrix that we are utilizing. Explicitly, it has the form:
\[
\Gamma^{i_1 i_2}_{j_1 j_2} =\frac{1}{2!}\langle \Psi | a_{i_1}^\dagger
a_{i_2}^\dagger a_{j_2}a_{j_1}  |  \Psi  \rangle.
\]
When dealing with the 1-body terms (present in most Hamiltonians of interest and several properties operators), the second-order reduced density matrix reduces to the first-order reduced density matrix defined as:
\[
\gamma^{i}_{j} =\langle \Psi |a_i^\dagger a_j  | \Psi\rangle,
\]
where $a^\dagger$ and $a$ denote the creation and annihilation operators,
respectively, and $\Psi$ is the $N$-particle antisymmetric wave-function.
Note that it is usually denoted by $\gamma^{i}_{j}$ instead of by $\Gamma^{i}_{j}$.

The second-order reduced density matrix has seen renewed interest for computing dynamical
properties of a quantum mechanical system governed by the electronic Hamiltonian.
When this descriptor was first introduced as a descriptor for electronic structure it was met with enthusiasm \cite{OldRDM}. Unfortunately, when the RDM method was applied to nuclear systems like 
${}^{24}\rm Mg$, ${}^{28}\rm Si$, the energies were found to be far below the expected value \cite{MihailovicRosina}. This is because the second-order reduced density matrix that resulted from these calculations did not originate from any wave-function \cite{Coleman63}! Every reduced density matrix of interest must result from some wave-function (this wave-function is known as the ancestor wave-function).  The problem of reduced density matrices not arising from ancestor wave-functions is what Coleman \cite{Coleman63} coined the $N$-representability problem.  Currently the necessary and sufficient conditions that guarantee $N$-representability are not known in any practical form \cite{Liu07}. Fortunately, several necessary conditions are known.  Using only the $P$, $Q$, $G$, $T1$, and $T2^\prime$ (necessary) conditions have been shown to reliably obtain chemical accuracy \cite{Zhao04,Nakata08}.

\subsection{$N$-representability conditions}

$N$-representability is the necessary and sufficient conditions that a density matrix originates from some (ancestor) wave-function \cite{Coleman63}. For the first-order density matrix to be $N$-representable its eigenvalues should lie in the closed interval $[0, 1]$ \cite{Kuhn60,Coleman63}. Since we know these conditions in an implementable form and due to Gilbert's theorem \cite{Gilbert75}, one can construct a method using only the 1-RDM. This method is sometimes referred as the density-matrix functional theory (DMFT) method \cite{DMFT}. The $N$-representability conditions are not limited to density matrices. The $N$-representability conditions for the electron density are known \cite{Gilbert75}. For the wave-function itself they are very simple, simply ensure the basis functions are square integrable and antisymmetric (change sign) with respect to the interchange of any two electron coordinates (Pauli principle).

Unfortunately, the second-order density matrix $N$-representability conditions are not known in any useful form (i.e. an uncountable set of conditions) \cite{Garrod64,Liu07}. However, many necessary conditions are known. 
Some trivial conditions are trace conditions;
\begin{equation}
 \sum_i \gamma^{i}_{i} = N, \,\,\,\,\, \sum_{ij} \Gamma^{ij}_{ij} = N(N-1)/2, \label{trace1}
\end{equation}
and
\begin{equation}
 \gamma^{i}_{j} = \frac{(N-1)}{2} \sum_{k} \Gamma^{ik}_{jk}.\label{trace2}
\end{equation}
An incomplete list of necessary conditions alone are not enough to guarantee $N$-representability. However, using them within the RDM method gives strict lower bounds to the energy. The general strategy within the RDM method is to choose necessary conditions that are easily implementable, computationally inexpensive, and result in accurate energies.  Of course, the more necessary conditions used the better the answer (though how much the energy is improved depends on the system being investigated and the necessary condition being used).  The most commonly utilized conditions used within the RDM method are positive-semidefinite type of $N$-representability conditions; the $P$, $Q$ \cite{Coleman63} and $G$ conditions \cite{Garrod64}. 

The $P$-condition is formulated by starting from the simple fact that if $A$ is an arbitrary one-particle operator, then the expectation value of $A^\dagger A$ should be non-negative,
\[
 \langle A^\dagger A \rangle = {\rm Tr} (A^\dagger A)\Gamma \geq 0.
\]
If we restrict $A$ to $A = \sum c_{ij} a_i a_j$, for an arbitrary set of real numbers $c_{ij}$, then
\[
 \langle A^\dagger A \rangle = \sum_{ijkl} c_{ij} c_{kl} \Gamma^{ij}_{kl} \geq 0.
\]
should be satisfied. Therefore, $\Gamma$ should be positive semidefinite. Explicitly, the $P$-condition is:
\[
 \Gamma^{i_1 i_2}_{j_1 j_2} = \langle \Psi | a^\dagger_{i_1}  a^\dagger_{i_2} a_{j_2} a_{j_1} | \Psi \rangle \succeq O,
\]
where we used $\X \succeq \O$ to indicate it is positive semidefinite. If we restrict $A$ to $A = \sum c_{ij} a^\dagger_i a^\dagger_j$, then likewise, the $Q$-condition \cite{Coleman63} is explicitly: 
\[
 Q^{i_1 i_2}_{j_1 j_2} = \langle \Psi | a_{i_1}  a_{i_2} a^\dagger_{j_2} a^\dagger_{j_1} | \Psi \rangle \succeq \O.
\]
The $Q$ matrix should also be positive semidefinite. The $G$-condition can be derived if we restrict $A$ to $A = \sum c_{ij} a^\dagger_i a_j$. The $T1$- and $T2$-conditions are derived in a somewhat more involved way \cite{Zhao04}. If we take $A$ as $A=\sum_{ijk} c_{ijk} c_{lmn} (a^\dagger_i a^\dagger_j a^\dagger_k a_n a_m a_{\ell}
+a_n a_m a_{\ell} a^\dagger_i a^\dagger_j a^\dagger_k )$, which is a positive semidefinite three-particle operator, but it cancels out the genuine three-particle part. Thus we can evaluate it using the second-order reduced density matrix. This is the $T1$ condition. In the same way, if we take $A=\sum_{ijk} c_{ijk} c_{lmn}  (a^\dagger_i a^\dagger_j a_k a^\dagger_n a_m a_{\ell} + a^\dagger_n a_m a_{\ell} a^\dagger_i a^\dagger_j a_k )$, which is also a three-particle operator, and then cancels the three-particle part again, then we have the $T2$-condition. Usually, the $T2'$ condition is used instead of the $T2$-condition. The $T2^\prime$-condition is an enhancement of the $T2$-condition that arises from the addition of the one-particle operator \cite{Braams07}.

\subsection{The RDM Method}
As stated in the previous section, the necessary and sufficient conditions for $N$-representability are not known in any useful form for the second-order reduced density matrix. However, many necessary conditions are known. Selecting necessary conditions for $N$-representability would be how we find the set  ${\tilde {\cal E}}_N$, the set of approximately (necessary) $N$-representable second-order reduced density matrices. Then the RDM method for the ground state is the minimization of the total energy subject to: 
\begin{eqnarray}
E_g & = & \min_{{\tilde {\cal E}}_N \ni \Gamma} \left \{ {\rm Tr} H \Gamma \right \}, \nonumber \\
    & = & \min_{{\tilde {\cal E}}_N \ni \Gamma} \left \{ \sum_{ij} v^i_j \gamma^i_j + \sum_{i_1 i_2 j_1 j_2} w^{i_1 i_2}_{j_1 j_2} \Gamma^{i_1 i_2}_{j_1 j_2} \right \}, \nonumber
\end{eqnarray}
where we used the definition of the (electronic) Hamiltonian as follows:
\[
H = \sum_{ij} v^i_j a^\dagger_i a_j+\frac{1}{2} \sum_{i_1 i_2 j_1 j_2} w^{i_1 i_2}_{j_1 j_2} a^\dagger_{i_1} a^\dagger_{i_2} a_{j_2} a_{j_1}.
\]
Finding and implementing conditions that expedite calculations is the strategy for increasing the accuracy of the RDM method. One can formulate a method for each of the $M$-th order reduced density matrices if the appropriate boundary conditions are included. All of the matrices used in the RDM method are positive semidefinite. As a result of this, a semidefinite optimization program is used \cite{Nakata01,Zhao04}. 

As mentioned in the previous section, an interesting difference between the variational methods and the RDM method are we always obtain {\it lower bounds} to the exact energy whereas the variational methods gives the upper bounds \cite{DM1960,Coleman63}.

\subsection{Semidefinite Programming}
Semidefinite programming (SDP) is a convex optimization problem,
a smooth generalization of the linear programming. The linear programming optimizes a linear functional of a non-negative vector whereas the SDP optimizes
a linear functional of a non-negative, ({\it i.e.}, positive semidefinite symmetric) matrix.
An SDP problem can be defined as:
\begin{eqnarray*}
 (P) & & \min : \Sigma_{k=1}^m c_k x_k \\
 & & \mbox{s.t. : } \X = \Sigma_{k=1}^m \F_k x_k - \F_0, \X \succeq \O,
\end{eqnarray*}
where the symbol $\SMAT^n$ is the space of $n \times n $ symmetric
matrices. We use $\X \succeq \O (\X \succ \O)$ to indicate 
$\X \in \SMAT^n $ is positive semidefinite (positive definite,
respectively). The Lagrangian dual $(D)$ of the problem $(P)$ can be derived as: 
\begin{eqnarray*}
 (D) & & \max : \F_0 \bullet \Y \\
 & & \mbox{s.t. : } \F_k \bullet \Y = c_k \ (k=1,\ldots,m) \\
 & & \Y \succeq \O.
\end{eqnarray*}
Here we used the Hilbert-Schmidt inner product $\U\bullet\V$ defined as $\sum_{i=1}^n \sum_{j=1}^n U_{ij}V_{ij}$ for $\U$ and $\V$ in $\SMAT^n$.

To solve the $(P)$ formulation or the $(D)$ formulation, we employ the primal-dual path-following interior-point method (PDIPM). This algorithm solves these two problems, $(P)$ and $(D)$, simultaneously in polynomial time. It is a widely accepted method because there are also many efficient implementations \cite{PDIPM}. One of the most efficient
implementation is SDPARA developed Fujisawa {\it et al.} \cite{SDPA}. We used this program to solve large scale semidefinite programming problems arising from condensed matter physics.

Here we briefly sketch a framework of the PDIPM:
\begin{description}
 \item[{\small Step} 0:] Choose an initial point $\x^0,\X^0,\Y^0$ with
            $\X^0 \succ \O, \Y^0 \succ \O$. Set $h=0$ and
            choose the parameter $\gamma \in (0,1)$.
 \item[{\small Step} 1:] Evaluate the Shur Complement Matrix $\B \in \SMAT^{n}$ by the formula
            \begin{eqnarray}
             B_{ij} = ((\X^h)^{-1} \F_i  \Y^h) \bullet \F_j.
              \label{eq:SCM}
            \end{eqnarray}
 \item[{\small Step} 2:] Solve the linear equation $\B d\x = \r$.
            Using its solution $d\x$,
            compute $d\X,d\Y$ and
            obtain the search direction
            $(d\x,d\X,d\Y)$.
 \item[{\small Step} 3:] Compute the maximum step length $\alpha$ to 
            keep the positive semidefiniteness;
            $\alpha = \max \{\alpha \in [0,1]:
            \X^h + \alpha d\X \succeq \O,
            \Y^h + \alpha d\Y \succeq \O \}$.
 \item[{\small Step} 4:] Update the current point by
            $ (\x^{h+1},\X^{h+1},\Y^{h+1}) =
            (\x^{h},\X^{h},\Y^{h})
            + \gamma \alpha (d\x,d\X,d\Y)$.
 \item[{\small Step} 5:] If $(\x^{h+1},\X^{h+1},\Y^{h+1})$ satisfies
            the stopping criteria, output it as a solution.
            Otherwise, set $h = h+1$ and return to Step 1.
\end{description}

Since basically, all the $N$-representability conditions can be written as linear inequalities \cite{Garrod64}, the RDM method with any kind of $N$-representability condition can be formulated as semidefinite programming problem.

There are two ways of formulating the RDM method as a standard type semidefinite programming problem.
One is the primal formulation \cite{Nakata01} and the other is the dual formulation \cite{Zhao04}. The primal formulation
is somewhat more involved than the dual formulation, however, the number of variables is reduced considerably. In either case, the formulation would result in large size SDPs. 

\subsection{The Hubbard Model}
The Hubbard model is a simple lattice model that was formulated to model strong correlation \cite{HUBBARD63}. It was used initially to show the behavior of $d$ electrons. In its original application the Hubbard model was used to describe electrons in solids.  It is now used for predicting superconductivity, particularly, it is used to describe the transition between conducting and insulating bands \cite{onedhubbard}. The Hubbard Hamilton for fermions has the form:
\[
H = -t \sum_{\langle i,j \rangle}^{L} \sum_{\sigma=\uparrow, \downarrow }
a^\dagger_{i,\sigma} a_{j,\sigma} + U \sum_{j=1}^{L} a^\dagger_{j,\uparrow} a_{j,\uparrow}
a^\dagger_{j,\downarrow} a_{j,\downarrow}
\]
where $U$ and $t$ are real parameters, $L$ is the number of sites on the lattice, and
$\langle i,j \rangle$ means summing over every $i$-th and $j$-th sites that is the nearest neighbors.
The Hamiltonian is made up of two terms. The first term is the kinetic term allowing for tunneling. It is also referred to as the hopping term.  The second term is the on-sight repulsion term.  The tight binding model is a special case of the Hubbard model. It arises when the on-site repulsion term is neglected ($U=0$).   It has this name because it describes tightly bonded electrons in solids. There are limited interactions of each particle with its neighbors.  This is similar to orbitals within a free atom in the standard linear combination of atomic orbitals (LCAO) model.  Alternatively, if $t=0$ we obtain the one-site model. As the name implies, all the sites are independent of each other (the neighbors no longer matter). Each site may contain $0$ electrons, $1$ alpha electron, $1$ beta electron, or $1$ alpha and $1$ beta electron. When $t \neq 0$ then only the ratio of $U/t$ matters \cite{onedhubbard}. It is a model that is easily adaptable to using the RDM method.

\section{Results and Discussion}
We targeted the repulsive 2-dimensional $4\times 4$ Hubbard model with periodic boundary condition, and using hopping term that is non-zero for nearest neighbors and zero otherwise. The number of electrons in the system is 16 (half-filled)
and the total spin of the system being $S^2=0$. Then, we solved for energies using Hamiltonians with $U/t$ values ranging from $0.01$
to $100$. Note that in the bipartite lattice case, there is a symmetry in $U/t$; changing the sign of $t$ does not alter the
 physics (in that only the ratio of $U/t$ matters), thus the same 2-RDM is obtained since the gauge transformation on the one sub lattice $A$ is:
\[
\{a_{i, \sigma}^{\dagger} \}_{A} \to \{-a_{i, \sigma}^{\dag} \}_{A}
\]
can change $t$ to $-t$. Thus, in this paper, we fixed to $t=1$. We employed the $P$, $Q$, $G$, $T1$, and $T2^\prime$ as $N$-representability
conditions and three types of combinations with trivial $N$-representability conditions (eqs. \ref{trace1} and \ref{trace2}) are examined; (i) the $P$, $Q$ and $G$ conditions, (ii) the $P$, $Q$, $G$, and $T1$ 
, (iii) the $P$, $Q$, $G$, $T1$, and $T2^\prime$ conditions. To compare the accuracy of each calculation we also performed the Exact Diagonalization 
method using ALPS \cite{ALPS}.

In Table \ref{machine} we show the details of the cluster machine used to solve the large SDPs of the 2-dimensional Hubbard model,
we show the size of the problem in Table \ref{sizeofsdps}, and we show typical time elapsed to solve the problem in Table \ref{elapsedtime}. As we see, the number of constraints do not change in the dual formulation \cite{Zhao04}, however the number of variable of the matrix becomes larger when we add the $T2^\prime$ conditions; the order of size is the same as in the third-order matrices. No simple collapse can be found as the $T1$ condition's case. The elapsed time shows that to solve using the $P$, $Q$, and $G$-conditions that it took approximately $1000$ to $2000$ seconds, to solve using the $P$, $Q$, $G$, and $T1$-conditions it took approximately $4000$ to $5000$ seconds,
and to solve using the $P$, $Q$, $G$, $T1$ and $T2^\prime$-conditions it took approximately $20,000$ to $30,000$ seconds.
Since the convergence criteria is well understood in the PDIPM, the number of iterations required to converge is dependent
of the size of the problem, and not on the form of the Hamiltonian. This is another feature of the RDM method when solved using the PDIPM.

We have omitted the results here, but the numerical quality of the solutions (the convergence) is approximately eight significant decimal digits. Theoretically, the primal-dual gap, primal and dual feasibility of the SDP $(P)$ and $(D)$ should be zero or numerically they should be sufficiently small (as determined by convergence criteria) \cite{SDPA}. These values
for our solutions are very small in our results; typical values for the 
primal-dual gap are:  $1.0\times 10^{-10}$ to $1.0\times 10^{-15}$,
and for the primal and dual feasibility are: $1.0\times 10^{-7}$ to $1.0\times 10^{-9}$.

In Table \ref{energy}, we show the total energy of various values of $U/t$ by (i) the ED method, (ii) by the $P$, $Q$, and $G$, (iii) by the $P$, $Q$, $G$, and $T1$, and (iv) by the $P$, $Q$, $G$, $T1$, and $T2^\prime$ conditions. The difference in the energies from the ED method, $\Delta E_{PQG} = E_{PQG} - E_{ED} $, $\Delta E_{PQGT1} = E_{PQGT1} - E_{ED} $, and $\Delta E_{PQGT1T2^\prime} = E_{PQGT1} - E_{ED} $ are shown.
Since adding $N$-representability conditions would result in increasing (and thus improving) the energy, therefore, $E_{PQG} \leq E_{PQGT1} \leq E_{PQGT1T2^\prime} \leq E_{ED}$ always holds theoretically, and we confirm this result numerically.

In the weak coupling limit $|U/t| << 1$, same as in the one-dimensional case, the total energies calculated by the RDM method almost coincide with the ED method; at $U/t=0.01$, the total energies from the RDM method is almost the same as the ED method.
As the coupling becomes larger, the total energy decreases rather quickly for $E_{PQG}$, especially when $U/t$ is larger than 1.
The worst energies are obtained when $4 < U/t < 10$. By adding the $T1$ condition, the absolute difference in energy reduces by a factor of two for most values of $U/t$. The most accurate results were obtained by when using the $P$, $Q$, $G$, $T1$ and $T2^\prime$ as the $N$-representability conditions. The largest error was obtained at $U/t=8$ ($-2.39 \times 10^{-1}$) or, equivalently, $-1.8 \times 10^{-2}$ per site.
This is still a very good energy \cite{Dagotto}. In the range where the worst energies were obtained, $4 < U/t < 10$, and is an important range because
in this parameter region quantum phase transition from metal (tight-binding one) to Mott-insulator
phase (and to the N\'e{}el phase) typically occurs. For this reason, the calculation of intermediate coupling is difficult \cite{onedhubbard}.

We cannot show the results for $U/t>1000$, since we faced the numerical difficulty in this region. In the high correlation limit $(|U/t|\rightarrow \infty$, all the states are nearly degenerated. In this case, SDP problems also become degenerated. This results numerical instability and we need to perform high precision calculations \cite{Nakata08}.

It is not clear what happens if we add more sites and electrons. The RDM method with $P$, $Q$, and $G$ is not
size-extensive; the total energy should scale if the size of the system is scaled \cite{Nakata09}. In molecular cases, reasonable bounds were found, but there is no guarantee to have a lower bound on the energy of the unit lattice. Size-extensively may be recovered when we 
add unitary invariant $N$-representability condition \cite{Nakata12}.

\begin{table} 
\caption{The following machines and SDP solver have been used in calculating the ground state energy of the Hubbard model.} 
\label{machine}
\begin{tabular}{c|c} \hline
Cores & 16 Nodes, 32 CPUs, 128 CPU cores \\ \hline
CPU & Intel Xeon 5460 3.16GHz (quad cores) x 2 / node \\\hline
Memory & 48GB / node \\\hline
NIC & GbE x 2 and Myrinet-10G x 1 / node \\\hline
OS & CentOS 5.8 for x86\_64 \\ \hline 
SDP Solver & SDPARA 7.3.2 RC2 \\ \hline 
\end{tabular}
\end{table}

\begin{table} 
\caption{Size of the SDPs for the $4\times 4$ Hubbard model using various $N$-representability conditions. The number of
constraints, the size of variable matrix, and the number of block matrices in the variable matrix and the maximum size
of the block in the variable matrix are shown.} 
\label{sizeofsdps}
\begin{tabular}{c|c|c|c|c} \hline
$N$-representability &  constraints & size of variables matrix & \# of blocks & maximum block\\ \hline
$P$, $Q$, $G$ & 47688 &  2634 &  14 & 512 \\ \hline
$P$, $Q$, $G$, $T1$ & 47688  & 7594 & 18 & 1920 \\ \hline
$P$, $Q$, $G$, $T1$, $T2^\prime$ & 47688 & 23498 & 22 & 6032 \\ \hline
\end{tabular}
\end{table}

\begin{table} 
\caption{Typical elapsed times to solve the SDP problems. }
\label{elapsedtime}
\begin{tabular}{c|c} \hline
$N$-representability &  Time (s)\\ \hline
$P$, $Q$, $G$ & $1000 \sim 2000$ \\
  $P$, $Q$, $G$, $T1$ & $4000 \sim 5000$ \\
$P$, $Q$, $G$, $T1$, $T2^\prime$ & $20,000 \sim 30,000$ \\ \hline
\end{tabular}
\end{table}

\begin{table}
\caption{The total energy of the ground state of the 2-dimensional Hubbard model with various $U/t$ and various methods: $E_{ED}$ is by the ED method, $E_{PQG}$ is by the RDM method using the $P$, $Q$, and $G$-conditions,
, $E_{PQGT1}$ is by the RDM method using the $P$, $Q$, $G$, $T1$-conditions,
, $E_{PQGT1T2^\prime}$ is by the RDM method using the $P$, $Q$, $G$, $T1$ and $T2^\prime$-conditions. $\Delta E_{PQG} = E_{ED} - E_{PQG}$,
and so on, and these $\Delta E_{PQG}, \Delta E_{PQGT1}$ and  $\Delta E_{PQGT1T2^\prime}$ should be negative.
 }
\label{energy}
\resizebox{17cm}{!}{
\begin{tabular}{r|l|l|l|l|l|l|l} \hline
$U/t$   & $E_{ED}$ & $E_{PQG}$ &  $E_{PQGT1}$ &  $E_{PQGT1T2^\prime}$ & $ \Delta E_{PQG}$  & $ \Delta E_{PQGT1}$  &  $ \Delta E_{PQGT1T2^\prime}$ \\ \hline
$  0.01 $ & $-23.9656 $ & $-23.9657$ & $-23.9657$ & $-23.9657$   & $-2.39\times 10^{-5}$ & $-1.59\times 10^{-5}$ & $ -1\times 10^{-7} $ \\
$  0.1  $ & $-23.6587 $ & $-23.6606$ & $-23.6599$ & $-23.6587$ & $-1.98\times 10^{-3}$ & $-1.22\times 10^{-3}$ & $ - 1\times 10^{-5} $\\
$  0.2  $ & $-23.3221 $ & $-23.3298$ & $-23.3268$ & $-23.3221$ & $-7.74\times 10^{-3}$ & $-4.74\times 10^{-3}$ & $ - 1\times 10^{-5} $\\
$  0.5  $ & $-22.3402 $ & $-22.3858$ & $-22.3682$ & $-22.3411$ & $-4.55\times 10^{-2}$ & $-2.79\times 10^{-2}$ & $-8.64 \times 10^{-4} $\\
$  0.8  $ & $-21.3991 $ & $-21.5090$ & $-21.4666$ & $-21.4024$ & $-1.10\times 10^{-1}$  & $-6.75\times 10^{-2}$ & $-3.31 \times 10^{-3} $\\
$  1    $ & $-20.7936 $ & $-20.9584$ & $-20.8953$ & $-20.7998$ & $-1.65\times 10^{-1}$ & $-1.02\times 10^{-1}$ & $-6.13 \times 10^{-3} $\\
$  2    $ & $-18.0176 $ & $-18.5478$ & $-18.3522$ & $-18.0535$ & $-5.30\times 10^{-1}$ & $-3.35\times 10^{-1}$ & $-3.60 \times 10^{-2} $\\
$  3    $ & $-15.6367 $ & $-16.5790$ & $-16.2473$ & $-15.7243$ & $-9.42\times 10^{-1}$ & $-6.11\times 10^{-1}$ & $-8.77 \times 10^{-2} $\\
$  4    $ & $-13.6219 $ & $-14.9454$ & $-14.4941$ & $-13.7711$ & $-1.32$ & $-8.72\times 10^{-1} $ & $-1.49 \times 10^{-1} $\\
$  5    $ & $-11.9405 $ & $-13.5745$ & $-13.0214$ & $-12.1479$ & $-1.63$ & $-1.08$ & $-2.07 \times 10^{-1} $\\
$  6    $ & $-10.5522 $ & $-12.4134$ & $-11.7728$ & $-10.8045$ & $-1.86$ & $-1.22$ & $-2.52 \times 10^{-1} $\\
$  7    $ & $-9.41048 $ & $-11.4208$ & $-10.7033$ & $-9.69084$ & $-2.01$ & $-1.29$ & $-2.80 \times 10^{-1} $\\
$  8    $ & $-8.46888 $ & $-10.5641$ & $-9.77809$ & $-8.76192$ & $-2.10$ & $-1.31$ & $-2.93 \times 10^{-1} $\\
$  9    $ & $-7.68624 $ & $-9.81887$ & $-8.97982$ & $-7.98034$ & $-2.13$ & $-1.29$ & $-2.94 \times 10^{-1} $\\
$  10   $ & $-7.02900 $ & $-9.16556$ & $-8.28788$ & $-7.31630$ & $-2.14$ & $-1.26$ & $-2.87 \times 10^{-1} $\\
$  100  $ & $-7.68192 \times 10^{-1}$ & $-1.21923$ & $-8.75302 \times 10^{-1} $ & $-7.83706 \times 10^{-1}$ & $-4.51\times 10^{-1}$ & $-1.07 \times 10^{-1} $ & $-1.55 \times 10^{-2}$ \\ \hline
\end{tabular}
}
\end{table}

\section{Conclusion}
The $P$, $Q$, $G$, $T1$ and $T2^\prime$ conditions were found to give reliable accuracy for the strongly
correlated electron case as espoused by the two-dimensional Hubbard model. The most difficult case for 
these conditions to resolve was the intermediate case where $4 \leq U/t \leq 10$. The accuracy was more than satisfactory when using the $P$, $Q$, and $G$ conditions. The deviation of the energy is $0.1$ unit for one lattice in this region.
By adding the $T1$ condition, the deviations from the exact energies were
improved by approximately a factor two from using only the $P$, $Q$ and $G$-conditions. 
Prominent results were obtained further adding $T2^\prime$ condition.  The deviations from the exact energies are approximately
 $0.01$ unit for intermediate couplings. 
If one could formulate an $N$-representability condition for a tightly bound electron with some small
probability of mobility this would increase the accuracy further. Nevertheless, the results in these cases are satisfactory and better in all other cases.
Since solving semidefinite programming problems from condensed matter physics becomes extremely large, then reducing the size by employing symmetry of the system is very important. The size of systems we calculated are too small to be conclusive and still not yet competitive
with other methods. Anyway, it is a very good challenge for the optimization community and high performance computing community. The reduced-density matrix method has been a promising method of quantum chemistry, however, in this paper, we showed
that this method is also promising for condensed matter physics where the electron correlation are very strong.

\subsection*{Acknowledgments}
We devote this paper of A.~J.~Coleman who contributed and motivated the researchers to the reduced density matrix method. J.~S.~M~A. is grateful to Prof. Coleman for all of the sound academic and career advice he received from him.
This research was partially supported by the Japan Science and Technology Agency (JST) Core Research of Evolutionary Science and Technology (CREST) research project. M.~N. was supported by the Special Postdoctoral Researchers' Program of RIKEN, and the study is partially supported by Grant-in-Aid for Scientific Research (B) 21300017. J.~S.~M~A. was supported by a postdoctoral fellowship from Japan Society for the Promotion of Science for foreign researchers. R.~I. was supported by the Strategic Programs for Innovative Research (SPIRE), MEXT, and the Computational Materials Science Initiative (CMSI), Japan, and M.~Y. was partially supported by Grant-in-Aid for Young Scientists (B) 24710161.

\end{document}